\documentclass[aps,prd,onecolumn,preprintnumbers,groupedaddress,showpacs,nofootinbib,amssymb]{revtex4}
\usepackage{graphicx,bm}
\usepackage{amsmath}
\usepackage{amssymb}
\usepackage{amsfonts}

\begin{document}

\def\pp{{\, \mid \hskip -1.5mm =}}
\def\cL{{\cal L}}
\def\be{\begin{equation}}
\def\ee{\end{equation}}
\def\bea{\begin{eqnarray}}
\def\eea{\end{eqnarray}}
\def\tr{\mathrm{tr}\, }
\def\nn{\nonumber \\}
\def\e{\mathrm{e}}

\newcommand{\f}[2]{
\frac{#1}{#2}
}

\newcommand{\tf}[2]{
\tfrac{#1}{#2}
}

\newcommand{\R}{
\right
}

\newcommand{\Lt}{
\left
}

\newcommand{\p}{
\partial
}

\newcommand{\n}{
\nonumber
}

\tolerance=5000

\title{Spinor with Schr\" odinger symmetry and 
non-relativistic supersymmetry}

\author{Hiroshi Yoda$^1$, Shin'ichi Nojiri$^{1,2}$}

\affiliation{
${}^1$Department of Physics, Nagoya University,~Nagoya 464-8602, Japan~\\
${}^2$Kobayashi-Maskawa Institute for the Origin of Particles and the
Universe, Nagoya University, Nagoya 464-8602, Japan}

%\date{\today}

\begin{abstract}

We construct the $d$ dimensional ``half'' Schr\" odinger equation, which is a kind 
of the root of the Schr\" odinger equation, from the $d+1$ dimensional free 
Dirac equation. The solution of the ``half'' Schr\" odinger equation 
also satisfies the usual free Schr\" odinger equation. 
We also find that the explicit transformation laws of the Schr\" odinger 
and the half Schr\" odinger fields under the Schr\" odinger symmetry 
transformation are derived by starting from the Klein-Gordon equation 
and the Dirac equation in $d+1$ dimensions. 
We derive the $3$ and $4$ dimensional super-Schr\" odinger algebra 
from the superconformal algebra in $4$ and $5$ dimensions. The algebra is realized 
by introducing two complex scalar and one (complex) spinor fields and the 
explicit transformation properties have been found. 

\end{abstract}


\maketitle

\section{Introduction}

Motivated with the original AdS/CFT 
correspondence \cite{Maldacena:1997re,Gubser:1998bc,Witten:1998qj}, 
non-relativistic conformal symmetry 
\cite{Son:2008ye,Balasubramanian:2008dm}, 
especially for the cold atom \cite{Nishida:2007pj} 
has attracted some attentions 
as a correspondence between the gravity in the anti-de Sitter (AdS) background and 
the non-relativistic conformal field 
theory \cite{Herzog:2008wg,Maldacena:2008wh,Adams,Goldberger:2008vg,Barbon:2008bg,Galilean conformal algebras and AdS/CFT}. 
The non-relativistic conformal symmetry is the Galilean conformal symmetry containing the Galilei symmetry and conformal symmetry.

The symmetry in the free Schr\" odinger equation \cite{Hagen:1972pd,Niederer:1972zz,Henkel:1993sg,Mehen:1999nd} , called as the Schr\" odinger symmetry,
 is  a kind of Galilean conformal symmetry.
The algebra of the Schr\" odinger symmetry in $d-1$ spacial dimensions is 
the subalgebra of $d+1$ dimensional conformal algebra, which is the algebra of 
the isometry of $d+2$ dimensional AdS space-time. 
Then the $d$ dimensional ($d-1$ spacial dimensional) free Schr\" odinger equation 
can be derived from the $d+1$ dimensional Klein-Gordon equation by choosing the wave 
function to be an eigenstate of the momentum of a light-cone direction.
In this paper, we define the ``half'' Schr\" odinger equation, which is a kind 
of the root of the Schr\" odinger equation, from the $d+1$ dimensional free 
Dirac equation\footnote{ 
%%%%%%%%%%%%%%%%
The Schr\" odinger symmetry for the spinor field has been investigated 
in \cite{Duval:1995fa}.
}.
%%%%%%%%%%%%%%%%%
We find the solution of the ``half'' Schr\" odinger equation 
also satisfies the free Schr\" odinger equation. 
Since the Klein-Gordon equation and the Dirac equation have conformal symmetry, 
the explicit transformation laws of the Schr\" odinger and the half Schr\" odinger 
fields under the Schr\" odinger symmetry transformation can be derived. 
Since there are scalar field(s) corresponding to the Klein-Gordon field(s) and 
spinor field(s) to the Dirac field(s), we may consider 
the supersymmetry \cite{Sakaguchi:2008rx,Sakaguchi:2008ku,Wen:2008hi,Sakaguchi:2009de}. 
%%%%%%%%%%%%%%%%%%%%
We should note that a non-relativistic supersymmetry for the vector and spinor 
has been considered in \cite{Horvathy:2010wj}. 
%%%%%%%%%%%%%%%%%%%%
Starting from the superconformal algebra in $4$ and $5$ dimensions, we derive the 
$3$ and $4$ dimensional super-Schr\" odinger algebra. The algebra is realized by introducing 
two complex scalar and one (complex) spinor fields. 

\section{Schr\" odinger symmetry from the Klein-Gordon and Dirac equations}

We now show that the $d$ dimensional free Schr\" odinger equation can be derived from 
$d+1$ dimensional d'Alembert equation (massless Klein-Gordon equation) by choosing the wave function to be an eigenstate of 
the momentum of a light-cone direction. 

We denote the coordinates in $d+1$ dimensional flat space-time by $\{ x^\mu \}$ 
$\left(\mu=0,\cdots d\right)$ and 
choose the metric tensor by $\eta^{\mu \nu}:= \mathrm{diag}(-1,1,\cdots ,1)$. 
We now define the light-cone coordinates by
\be
\label{SA1}
x^\pm :=\frac{1}{\sqrt{2}}(x^0 \pm x^d)\, .
\ee
Then in terms of the light-cone coordinates, the metric tensor has the following form:
\be
\label{SA2}
\eta ^{\mu \nu}=
\begin{pmatrix}
0 &-1 &&&& \\
 -1 & 0 &&&& \\
&& 1 && \\
&&& \ddots & \\
&&&& 1
\end{pmatrix} 
\ee
Now $\mu,\nu =+,-,i=1,\cdots d-1$. 
Then, for example, we have $x^\pm =-x_\mp$ and 
\be
\label{SA3}
p_\pm :=\frac{1}{\sqrt{2}}(p_0 \pm p_d)\, ,\quad 
p^\pm =-p_\mp \, .
\ee

Now the relativistic energy-momentum relationship for the massless particle 
$p^\mu p_\mu = -2p_+p_- +{\bf p}^2 =0$ can be rewritten if we write
\be
\label{SA4}
p_+ = \mathcal{M} \, , \quad p_-\equiv E \, ,
\ee
as
\be
\label{SA5}
E=\frac{1}{2\mathcal{M}}{\bf p}^2\, ,
\ee
which is nothing but the expression of the kinetic energy for the non-relativistic 
particle with mass $\mathcal{M}$ in $d$ dimensional space 
($d+1$ space-time dimensions). 

Eq.~(\ref{SA5}) may suggest that the massless Klein-Gordon equation
could be rewritten as the free Schr\" odinger equation. 
By using the light-cone coordinates, the Klein-Gordon equation has the following form:
\be
\label{SA6}
(-2\p _+ \p _- +\p _i \p _ i )\phi =0\, .
\ee
We now choose $\phi$ to be an eigenstate of $\p_+$ as
$\p _+\phi = -i\mathcal{M}\phi$ and we regard a light-cone coordinate 
$x^-$ as a ``time''-coordinate $x^- \equiv t$. 
Then the Klein-Gordon equation (\ref{SA6}) can be rewritten 
as the free Schr\" odinger equation:
\be
\label{SA7}
i\p _t \phi =-\frac{1}{2\mathcal{M}}\triangle \phi \, .
\ee
Since we assume $\phi$ to be an eigenstate of $\p_+$, $\phi$ has the following form:
\be
\label{SA8} 
\phi (x^+,x^-,x^i)=\e^{-i\mathcal{M}x^+}\tilde \phi (x^-,x^i) \, .
\ee
Later we consider the conformal transformation of the Schr\" odinger 
field $\phi$, where the transformation of $x^+$ can be interpreted 
as a phase of $\tilde\phi$. 

The free massless Dirac field also satisfies the Klein-Gordon equation. 
Then by rewriting the Dirac equation by choosing the Dirac field $\psi$ 
to be an eigenstate of $\p_+$ as
$\p _+\psi = -i\mathcal{M}\psi$ and regarding light-cone coordinate 
$x^-$ as a ``time''-coordinate $x^- \equiv t$, we obtain the Schr\" odinger 
version of the Dirac equation. 
We now use the following convention for the $\gamma$ matrices:
\bea
\label{SA9}
&& \{ \gamma ^\mu ,\gamma ^\nu \} =2\eta ^{\mu \nu} \, ,\quad
{\gamma ^\mu}^\dagger = \gamma ^0 \gamma ^\mu  \gamma ^0 \, ,\quad
\Sigma  ^ {\mu \nu}:=\frac{1}{4}[\gamma ^\mu ,\gamma ^\nu ] \, ,\nn
&& \gamma ^5:=i\gamma ^0\gamma ^1\gamma ^2\gamma ^3 \, ,\quad
\gamma ^\pm  :=\frac{1}{\sqrt{2}}(\gamma ^0 \pm \gamma ^d)
=-\frac{1}{\sqrt{2}}(\gamma _0 \mp \gamma _d)=:-\gamma _ \mp \, .
\eea
Then we obtain
\be
\label{SA10}
\{ \gamma ^+ ,\gamma ^- \} =-2 \, , \quad
\{ \gamma ^\pm  ,\gamma ^i \} =0 \quad \left( i=1,2,\cdots, d-1 \right)\, , \quad
(\gamma ^+)^2=(\gamma ^-)^2=0 \, .
\ee
Then by choosing the Dirac field $\psi$ to be an eigenstate of $\p_+$ as
$\p _+\psi = -i\mathcal{M}\psi$, the Dirac equation has the following form: 
\be
\label{SA11}
\left(i\gamma ^a \p _a +\mathcal{M}\gamma ^+ \right)\psi =0 \quad 
\left( a=-,i \right) \, ,
\ee
which tells $\psi$ satisfies the Schr\" odinger equation:
\be
\label{SA12}
(2i\mathcal{M}\p _t +\p _i \p _ i )\psi =0 \, .
\ee
Then Eq.~(\ref{SA11}) is the Schr\" odinger version of the Dirac equation, which we call 
the ``half Schr\" odinger equation''. 

The $d$-dimensional Schr\" odinger algebra is given as a subalgebra of 
the $d+1$ dimensional conformal algebra $SO(d+1,2)$, which is given by
\bea
\label{SA13} 
&& \left[M_{\mu \nu },M_{\rho \sigma } \right ] 
= i(\eta_{\mu \sigma } M_{\nu \rho } -\eta_{\mu \rho }  M_{\nu \sigma }
 -\eta_{\nu \sigma } M_{\mu \rho} +\eta _{\nu \rho }  M_{\mu \sigma } ) \, , \nn
&& \left[M_{\mu \nu},P_\rho \right ] = i(\eta_{\nu \rho } P_\mu -\eta _{\mu \rho } P_\nu )\, , \quad
\left [M_{\mu \nu},K_\rho \right] = i(\eta _{\nu \rho } K_\mu -\eta_{\mu \rho } K_\nu ) \, ,\nn
&& [D,P_\mu ] = -iP_\mu \, , \quad [D,K _\mu ] = iK_\mu \, , \quad
[K_\mu ,P_\nu ]=-2i(\eta _{\mu \nu } D+M_{\mu \nu}) \, .
\eea
Here $M _{\mu \nu }$ generates the Lorentz boost and the rotation and $P_\mu$ generates 
the translation. 
Furthermore the operator $D$ generates the dilatation 
\be
\label{SA14}
x^\mu \rightarrow \e^{-\lambda }x^\mu \, ,
\ee
and $K_\mu$ generates the special conformal transformation given by
\be
\label{SA15}
x^\mu \rightarrow \frac{x^\mu -c^\mu x^2}{1-2c^\mu x_\mu +c^2 x^2} \, .
\ee
The $d-1$ dimensional Schr\" odinger algebra $\mathfrak{sch}(d-1)$ 
is the subalgebra of the conformal algebra 
which does not change the eigenvalue $\mathcal{M}$ of $P_+$, that is, the algebra 
given by the generators commuting with $P_+$. 
\bea
\label{SA16}
&& [J _{ij},J _{kl}] =i(\delta  _{il} J_{jk} -\delta_{ik}  J_{jl} 
 -\delta _{jl} J_{ik} +\delta _{jk}  J_ {il} ) \, ,\nn
&& [J _{ij},H ] =0 \, ,\quad [J _{ij},E ] =0 \, ,\quad
[J _{ij},F ] =0 \, ,\nn
&& [P_i,P_j] =0 \, ,\quad
[P_i,H ] =0 \, ,\quad
[P_i,E ] =iP_i \, ,\quad
[P_i,F ] =0 \, ,\quad
[J _{ij},P _k] = i(\delta _{jk} P_i -\delta _{ik} P_j ) \, ,\nn
&& [G_i,G_j] =0\, , \quad
[G_i,H ] =iP_i \, ,\quad
[G_i,E ] =iG_i \, ,\quad
[G_i,F ] =0 \, ,\quad
[J _{ij},G_k] = i(\delta _{jk} G_i -\delta _{ik} G_j ) \, ,\nn
&& [P_i,G _j] =i\mathcal{M}\delta _{ij} \, ,\quad [E,H] =-2iH \, ,\quad
[E,F] =2iF \, ,\quad
[F,H] =iE \, .
\eea
Here 
\be
\label{SA17}
H=P_- \, ,\quad P_i=P_i \, ,\quad 
P_+=\mathcal{M} \, ,\quad 
J _{ij} =M_{ij} \, ,\quad 
G_i =M_{+i} \, ,\quad 
E=D+M_{-+} \, ,\quad 
F=\frac{1}{2}K_+ \, .
\ee
The generators $H$, $P_i$, $J_{ij}$, $G_i$, $E$, and $F$ generate 
the translation of time, the translation of the spacial coordinates, 
the spacial rotation, the Galilei boost, the dilatation, and the 
special conformal transformation, which are explicitly given as 
the following transformation of the coordinates:
\bea
\label{SA18}
&& H:\, t\rightarrow  t+a \, ,\quad 
P_i:\, x_i \rightarrow  x_i +b_i \, ,\quad
J_{ij}:\, x_i \rightarrow  R_{ij}x_j\ \left(R_{ij}\in SO\left(d-1\right)\right)  \, ,\quad 
G_i:\, x_i \rightarrow  x_i + v_i t \, ,\nn 
&& E:\, t\rightarrow  \lambda ^2 t \ \mbox{and} \ x_i \rightarrow \lambda x_i \, \quad 
F:\, t\rightarrow \frac{1}{1+ft}t\ \mbox{and} \ x_i \rightarrow \frac{1}{1+ft}x_i \, .
\eea
In (\ref{SA16}), since $\mathcal{M}$ can be regarded as a $c$ number, the commutator of $P_i$ and $G _j$ becomes a central extension. 
In fact, this algebra is one of the central extended of the Galilean conformal algebra which is constructed by nonrelativistic limit ($c\rightarrow \infty $) of  conformal algebra. 
\cite{Exotic Galilean Conformal Symmetry and its Dynamical Realisations}

As well known, the Klein-Gordon equation (\ref{SA6}) and the Dirac equation are invariant under the 
conformal transformation and therefore the Schr\" odinger equation (\ref{SA7}) 
and the half Schr\" odinger equation (\ref{SA11}) is invariant under the Schr\" odinger 
transformation. 
The invariance of the Klein-Gordon and the Dirac equations and the transformation law of the 
Klein-Gordon and the Dirac fields in general $D$ dimensions are explicitly given in the appendices. 

\section{Super-Schr\" odinger symmetry from superconformal symmetry}

Since we have found that both of the Klein-Gordon equation and the Dirac equation 
have the Schr\" odinger symmetry and found the explicit transformation laws of the 
Klein-Gordon and the Dirac fields, we now like to find the transformation 
connecting the Klein-Gordon field(s) and the Dirac field(s), 
which is the super-Schr\" odinger symmetry transformation. 

The supersymmetric extension of the conformal symmetry is known as the 
superconformal symmetry. 
In this section, for simplicity, we construct $(2+1)$ and  $(3+1)$ dimensional super-Schr\" odinger 
algebra starting with the $D=4$, $\mathcal N =1$ and $D=5$, $\mathcal N =2$ superconformal algebra.

\subsection{Construction of $(2+1)$ dimensional super-Schr\" odinger algebra form the reduction 
of $D=4$, $\mathcal N =1$ superconformal algebra}

$D=4$, $\mathcal N =1$ Superconformal algebra\cite{Properties of conformal supergravity}, which is given by, in addition to the conformal algebra $SO(4,2)$ (\ref{SA13}), 

\bea
\label{SA19}
&& \{ Q_\alpha ,  \bar{Q}_\beta \} = -\frac{1}{2}i \gamma^ \mu_{\alpha \beta}P_\mu \, , \quad
\{ S_\alpha ,  \bar{S}_\beta \} = \frac{1}{2}i \gamma^ \mu_{\alpha \beta}K_\mu \, , \quad 
\{ Q_\alpha ,  \bar{S}_\beta \} = -\frac{1}{2}i \delta_{\alpha \beta}D 
+\frac{1}{2}i \Sigma^{\mu \nu}_{\alpha \beta}M_{\mu \nu}
+\gamma^5_{\alpha \beta}A \, , \nn
&& [Q_\alpha , M_{\mu \nu}]=i (\Sigma_{\mu \nu}Q)_\alpha \, , \quad
[S_\alpha , M_{\mu \nu}]=i \left(\Sigma_{\mu \nu}S\right)_\alpha \, , \quad 
[Q_\alpha , D]=\frac{1}{2}iQ_\alpha \, ,\quad
[S_\alpha , D]=-\frac{1}{2}iS_\alpha \, , \nn
&& [Q_\alpha , K_\mu ]=-i(\gamma_\mu S)_\alpha \, ,\quad
[S_\alpha , P_\mu ]=i(\gamma_\mu Q)_\alpha \, ,\quad 
[Q_\alpha , A]=-\frac{3}{4}(\gamma ^5 Q)_\alpha \, ,\quad
[S_\alpha , A]=\frac{3}{4}\left(\gamma ^5 S\right)_\alpha \, .
\eea
In order to obtain the closed algebra, we define projection operators 
$\mathcal{P}$ and $\bar{\mathcal{P}}$ as follows
\be
\label{SA20}
\mathcal{P}:=-\frac{1}{2}\gamma ^- \gamma ^+ \, , \quad 
\bar{ \mathcal{P} }:=-\frac{1}{2}\gamma ^+ \gamma ^- \, ,
\ee
which satisfy the following equations:
\bea
\label{SA21}
&& 
\mathcal{P}+\bar{ \mathcal{P} }=-\frac{1}{2}\{ \gamma ^-,  \gamma ^+ \} =1 \, , \quad 
\mathcal{P}^2=\frac{1}{4}\gamma ^- \gamma ^+\gamma ^- \gamma ^+ 
=-\frac{1}{2}\gamma ^- \gamma ^+ = \mathcal{P} \, ,\nn
&&
 \mathcal{P}\gamma ^-=\gamma ^- \bar{ \mathcal{P} }=\gamma ^- \, , \quad 
\gamma ^+\mathcal{P}=\bar{ \mathcal{P} }\gamma ^-=\gamma ^+ \, ,\quad 
\mathcal{P}\gamma ^+=\gamma ^+ \bar{ \mathcal{P} }
=\gamma ^-\mathcal{P}=\bar{ \mathcal{P} }\gamma ^- =0 \, , \nn
&&
 \overline{ \eta \mathcal{P}}=-\frac{1}{2}\eta ^\dagger \mathcal{P}^\dagger 
=-\frac{1}{2}\bar{ \eta }\gamma ^+ \gamma ^- \, , \quad 
\bar{ \mathcal{P} }=-\mathcal{P}^\dagger :=-\frac{1}{2}\gamma ^+ \gamma ^- \, , \nn
&& 
\bar{\mathcal{P}}\gamma ^a \mathcal{P}= 0\quad (a =-,1,\dots ,d-1) \, , \quad
\bar{\mathcal{P}}\gamma ^+ \mathcal{P}= \gamma ^+ \, .
\eea
If we define 
\be
\label{SA22}
T_\alpha := \bar{ \mathcal{P}}_{\alpha \beta}S_\beta \, , \quad 
\bar{T}_\alpha :=\bar{S}_\beta  \mathcal{P}_{\beta \alpha} 
=\overline{\bar{ \mathcal{P}}_{\alpha \beta}S_\beta} \, ,
\ee
we obtain
\be
\label{SA23}
(\gamma ^+ T)_\alpha =(\bar{T} \gamma ^-)_\alpha =0 \, .
\ee
Then a closed algebra, which we call the super-Schr\" odinger 
algebra, is given by
\bea
\label{SA24}
&& \{ Q_\alpha ,  \bar{Q}_\beta \} = -\frac{1}{2}i \gamma^-_{\alpha \beta} H 
 - \frac{1}{2}i \gamma^i_{\alpha \beta}P_i - \frac{1}{2}i \gamma^+_{\alpha \beta}\mathcal{M} \, , \quad
\{ T_\alpha ,  \bar{T}_\beta \} = i\gamma_{\alpha \beta}^+ F \, , \nn
&& \{ Q_\alpha ,  \bar{T}_\beta \} = \frac{1}{4}i (\gamma^- \gamma^+ )_{\alpha \beta}E
 -\frac{1}{4}i (\Sigma_{ij} \gamma^- \gamma^+ )_{\alpha \beta}J _{ij} 
+\frac{1}{2}i (\gamma^+ \gamma_i )_{\alpha \beta}G_i 
 - \frac{1}{2}(\gamma^5\gamma^- \gamma^+)_{\alpha \beta}A \, , \nn
&& [Q_\alpha , J_{ij}]=i  (\Sigma_{ij}Q)_\alpha \, , \quad
[T_\alpha , J_{ij}]= i (\Sigma_{ij}T)_\alpha \, , \quad 
[Q_\alpha , G_i ] = -\frac{1}{2}i(\gamma^- \gamma_i Q)_\alpha \, ,\quad
[T_\alpha , G_i ]=0 \, , \nn
&& [Q_\alpha , E]=-\frac{1}{2}i(\gamma^- \gamma^+Q)_\alpha \, ,\quad
[T_\alpha , E]= - iT_\alpha \, ,\quad 
[T_\alpha , H ]= -i(\gamma^+Q)_ \alpha \, ,\quad
[T_\alpha , P_i ]= -\frac{1}{2}i(\gamma^+ \gamma^- \gamma _ i Q)_\alpha \, , \nn
&& [T_\alpha , F ]=0 \, , \quad 
[Q_\alpha ,F]=\frac{1}{2}i(\gamma^- T)_\alpha \, , \quad 
[Q_\alpha , A]= -\frac{3}{4}(\gamma ^5 Q)_\alpha \, , \quad
[T_\alpha , A]=\frac{3}{4}(\gamma ^5 T)_\alpha \, .
\eea

We now consider the following Lagrangian
\be
\label{SA25}
\mathcal{L}= i\bar\psi \gamma ^\mu \partial _\mu \psi 
 - \frac{1}{2}\sum _{i=i,2}\p _\mu \phi _i^* \p ^\mu \phi _i \, .
\ee 
The Lagrangian is invariant under the following superconformal transformation: 
\bea
\label{SA26}
&&
 \delta _Q\phi _1=-\frac{1}{2}i(\bar{\xi  }\psi ) \, , \quad
\delta _Q\phi _2=-\frac{1}{2}i(\bar{\xi  }\gamma ^5\psi ) \, , \quad 
\delta _Q\psi = \frac{1}{2}(i\gamma ^- \p _t +i\gamma ^i \p _i +M\gamma ^+)
(\xi  \phi _1-\gamma ^5 \xi  \phi _2) \, , \nn
&&
 \delta _T\phi _1=-\frac{1}{2}i(\bar{\eta} \bar{\mathcal{P}}(t\gamma ^+ -x_i\gamma _i ) \psi ) \, , \quad
\delta _T\phi _2=\frac{1}{2}i(\bar{\eta }\bar{\mathcal{P}}\gamma ^5(t\gamma ^+ -x_i\gamma _i ) \psi ) \, , \nn
&&
 \delta _T\psi =\frac{1}{2}i[(2 t \p _t +x_i \p_ i+2)+\gamma ^+\gamma ^i (t\p _ i-i\mathcal{M}x_i) 
 -\Sigma _{ij}(x_i \p _j-x_j \p _i)](\phi _1+\gamma ^5\phi _2) \mathcal{P}\eta \, , \nn
&&
\delta _A\phi _1=-\frac{1}{2}\theta \phi _2 \, ,\quad 
\delta _A\phi _2=\frac{1}{2}\theta \phi _1 \, ,\quad \delta _A\psi =-\frac{1}{4}\theta \gamma ^5\psi \, .
\eea
Here the $\xi $, $\eta $ and $\theta$ are parameters of the transformation. 
The transformation (\ref{SA26}) is closed on shell. 

As we are considering the field theory in 4 dimensions, we may define $\gamma^5$ 
and the chiral projection operators $\mathcal{P}_{L,R}$ by
\be
\label{SA27}
\gamma ^5:=i\gamma ^0\gamma ^1\gamma ^2\gamma ^3 \, ,\quad
\mathcal{P}_L:=\frac{1}{2}\left(1-\gamma ^5\right) \, ,\quad
\mathcal{P}_R:=\frac{1}{2}\left(1+\gamma ^5\right) \, .
\ee
We also define the left and right parts for the Dirac field and also the scalar field by 
\be
\label{SA28}
\psi _ L:=\mathcal{P}_L  \psi \, ,\quad 
\psi _ R:=\mathcal{P}_R  \psi \, ,\quad 
\phi _ L:=\phi _1-\phi_2 \, ,\quad 
\phi _ R:=\phi _1+\phi_2 \, .
\ee
Then the transformation laws (\ref{SA26}) can be rewritten as
\bea
\label{SA29}
&& 
\delta _{Q_L}\phi _L=-i(\bar{\xi }\psi _L) \, ,\quad
\delta _{Q_R}\phi _R=-i(\bar{\xi }\psi _R) \, ,\nn 
&&
\delta _{Q_R}\psi _L = \frac{1}{2}(i\gamma ^- \p _t +i\gamma _i \p _i + M\gamma ^+)\mathcal{P}_R \xi  \phi _L \, , \quad 
\delta _{Q_L}\psi _R = \frac{1}{2}(i\gamma ^- \p _t +i\gamma _i \p _i + M\gamma ^+)\mathcal{P}_L \xi  \phi _R \, , \nn
&&
\delta _{T_R}\phi _L=-ix(\bar{\eta} \bar{\mathcal{P}}(t\gamma ^+ -x_i\gamma _i ) \psi _L) \, ,\quad 
\delta _{T_L}\phi _R=-ix(\bar{\eta }\bar{\mathcal{P}}(t\gamma ^+ -x_i\gamma _i ) \psi _R) \, ,\nn
&&
\delta _{T_L}\psi _L=\frac{1}{2}i[(2 t \p _t + x_i \p _i +2)+\gamma ^+\gamma _i (t\p _i-i\mathcal{M}x_i) 
-\Sigma _{ij}(x_i \p _j - x_j \p _i)]\mathcal{P}\mathcal{P}_L\eta \phi _L \, ,\nn
&&
\delta _{T_R}\psi _R=\frac{1}{2}i[(2 t \p _t + x_i \p _i +2)+\gamma ^+\gamma ^i (t\p _i-i\mathcal{M}x_i) 
-\Sigma _{ij}(x_i \p _j - x_j \p _i)]\mathcal{P}\mathcal{P}_R\eta \phi _R \, , \nn
&&
\delta _{\tilde{A}}\phi _L= - \theta \phi _R \, , \quad 
\delta _{\tilde{A}}\phi _R = \theta \phi _L \, ,\quad 
\delta _{\tilde{A}}\psi _L = \frac{1}{2} \theta \psi _L \, ,\quad
\delta _{\tilde{A}}\psi _R= -\frac{1}{2} \theta \psi _R \, .
\eea
Here $\tilde{A}:=2A$. 
Except the transformation generated by $A$, the transformations are closed on the left and right parts, respectively. 

\subsection{Construction of $(3+1)$ dimensional super-Schr\" odinger algebra from the reduction 
of $D=5$, $\mathcal N =2$ superconformal algebra}

$D=5$, $\mathcal N =2$ Superconformal algebra
\cite{Superconformal Tensor Calculus in Five Dimensions}\cite{Superconformal N=2 D=5 matter with and without actions}
 is given by, 
\bea
\label{SA30}
&&
 \{ Q_{A\alpha} ,  \bar{Q}^B_ \beta \} = -\f{1}{2}i \delta _A^B\gamma ^ \mu _{\alpha \beta}P_ \mu  \, , \quad
\{ S_{A\alpha} ,  \bar{S}^B_ \beta \} = \f{1}{2}i \delta _A^B\gamma ^ \mu _{\alpha \beta}K_ \mu \, , \quad 
\{ Q_{A\alpha} ,  \bar{S}^B_ \beta \} = 
\delta _A^B\Lt( \f{1}{2}i \delta _{\alpha \beta}D-\f{1}{2}i \Sigma  ^ {\mu \nu} _{\alpha \beta}M _{\mu \nu} \R)
+\delta _{\alpha \beta}R_A^B  \, ,  \nn 
&& 
[Q_{A\alpha} , M _{\mu \nu}]=i  (\Sigma  _ {\mu \nu}Q_A)_\alpha \, , \quad
[S_{A\alpha} , M _{\mu \nu}]=i  (\Sigma  _ {\mu \nu}S_A)_\alpha \, , \quad 
[Q_{A\alpha} , D]=\f{1}{2}iQ_{A\alpha} \, ,\quad
[S_{A\alpha} , D]=-\f{1}{2}iS_{A\alpha} \, , \nn
&& 
[Q_{A\alpha} , K_\mu ]=-i(\gamma _ \mu S_A)_\alpha \, ,\quad
[S_{A\alpha} , P_\mu ]=i(\gamma _ \mu Q_A)_\alpha \, ,\nn
&&
[Q_{A\alpha} , R_{BC}]=\f{3}{4}(\varepsilon _ {AB}Q_{k\alpha}+\varepsilon _ {AC}Q_{j\alpha}) \, ,\quad
[S_{A\alpha} ,  R_{BC}]=\f{3}{4}(\varepsilon _ {AB}S_{k\alpha}+\varepsilon _ {AC}S_{j\alpha}) \, , \nn
&&
[R_{AB}, R_{CD}]=\f{3}{4}(\varepsilon _ {AC}R_{BD}+\varepsilon _ {BD}R_{AC}+ \varepsilon _ {AD}R_{BC}+\varepsilon _{BC}R_{AD}) \, 
\eea
and conformal algebra $SO(5,2)$ (\ref{SA13}).

Here  fermionic generators with  extra indices, $A,B,C...=1,2$ , is $SU(2)$ Majorana spinors 
and $R_{AB}=R_{BA}$ is $SU(2)$ R-symmetry generators.
These indices are raised and lowered with totally antisymmetric tensor $\varepsilon _{AB}$ 
and $\varepsilon ^{AB}$ ($\varepsilon _{12} = \varepsilon ^{12} =1$), as
\bea
\zeta ^A := \varepsilon ^{AB} \zeta _B,\quad \quad \zeta _A := \zeta ^B \varepsilon _{BA} \, ,
\eea
and  $SU(2)$ R invariant inner product is defined by contraction of raised and lowered indices,
\bea
\bar\zeta _1^A \zeta _{2A} = \varepsilon ^{AB} \bar\zeta _{1B}\zeta _{2A}=\varepsilon _{AB} \bar\zeta ^{1A}\zeta ^{2B} \, .
\eea
  
In the same way as in the $(2+1)$ dimensional case (\ref{SA22}), 
we reduce this algebra using by projection operators (\ref{SA20}), such as
\bea
&&\{ Q_{A\alpha} ,  \bar{Q}^B_ \beta \} =
-\f{1}{2}i \delta _A^B[\gamma ^ - _{\alpha \beta}H+\gamma _{i \alpha \beta}P_i+\gamma ^+ _{\alpha \beta}\mathcal{M} ] \, , \quad
\{ S_{A\alpha} ,  \bar{T}^B_ \beta \} = i\delta _A^B\gamma _{\alpha \beta}^+F  \, , \quad \nn
&&\{ Q_{A\alpha} ,  \bar{T}^B _ \beta \} = -\delta _A^B\Lt[ \f{1}{4}i (\gamma ^- \gamma ^+ )_{\alpha \beta}E 
-\f{1}{4}i (\Sigma  _ {ij}\gamma ^- \gamma ^+ ) _{\alpha \beta}J _{ij}
+\f{1}{2}i (\gamma ^+ \gamma _i )_{\alpha \beta}G_i  \R] 
-\f{1}{2}(\gamma ^- \gamma ^+) _{\alpha \beta}R_A^B  \, , \quad \nn
&&[Q_{A\alpha} , J _{ij}]=i  (\Sigma  _ {ij}Q_A )_\alpha \, , \quad
[T_{A\alpha} , J _{ij}]=i  (\Sigma  _ {ij}T_A )_\alpha \, , \quad
[Q_{A\alpha} , G_i ]=-\f{1}{2}i(\gamma ^- \gamma _i Q_A )_\alpha \, , \quad \nn
&&[T_{A\alpha} , G_i ]=0 \, , \quad
[Q_{A\alpha} , E]=-\f{1}{2}i(\gamma ^- \gamma ^+Q_A )_\alpha \, , \quad
[T_{A\alpha} , E]=-iT_{A\alpha} \, , \quad
[T_{A\alpha} , H ]=-i(\gamma ^+Q_A)_ \alpha \, , \quad\nn
&&[T_{A\alpha} , P_i ]=-\f{1}{2}i(\gamma ^+ \gamma ^- \gamma _ i Q_A)_\alpha \, , \quad 
[T_{A\alpha} , F ]=0 \, , \quad
[Q_{A\alpha} ,F]=\f{1}{2}i(\gamma ^ - T_A)_\alpha  \, , \quad \nn
&&[Q_{A\alpha} , R_{BC}]=\f{3}{4}(\varepsilon _ {AB}Q_{C\alpha}+\varepsilon _ {AC}Q_{B\alpha}) \, , \quad
[T_{A\alpha} ,  R_{BC}]=\f{3}{4}(\varepsilon _ {AB}T_{C\alpha}+\varepsilon _ {AC}T_{B\alpha})   \, , \quad \nn
&&[R_{AB}, R_{CD}]=\f{3}{4}(\varepsilon _ {AC}R_{BD}+\varepsilon _ {BD}R_{AC}+ \varepsilon _ {AD}R_{BC}+\varepsilon _{BC}R_{AD}) \, .
\eea

Thus, the on-shell transformation of hypermultiplet including two scalar fields $\phi ^A$ and one spinor field $\psi  _\alpha$ is  
\bea
&&\delta _Q\phi ^A=-\f{1}{2}i(\bar{\xi}^A\psi ) \, , \quad
\delta _Q\psi =(i\gamma ^- \p _t +i\gamma _i \p _i +\mathcal{M}\gamma ^+) \xi  _A \phi ^A  \, , \quad
\delta _T\phi ^A=-\f{1}{2}i(\bar\eta ^A \bar{\mathcal{P}}(t\gamma ^+ -x_i\gamma _i ) \psi ) \, , \quad \nn
&&\delta _T\psi =-i[(2 t \p _t+x_i \p _ i+3)+\gamma ^+\gamma _i (t\p_ i-i\mathcal{M}x_i)  
-\Sigma _{ij}(x_i \p _j-x_j \p _i)]\mathcal{P}\eta _A \phi ^A \, , \quad \nn
&&\delta _R \phi ^A=-\f{3}{2}\theta ^A _{\ B} \Phi ^B \, , \quad
\delta _R\psi =0 \, .
\eea
Here the $SU(2)$ Majorana spinor $\xi ^A,\eta ^A$ and $2\times 2$ (anti-)Hermitian traceless 
matrix $\theta ^A _{\ B}$ are parameters of the transformation.
In contrast to the 3 dimensional algebra (except the transformation generated by $A$), 
because of R-symmetry invariance of spinor field $\psi$, the transformations of two scalar fields $\phi^A$ are not closed. 

\section{Summary}

In this paper, we have found the ``half'' Schr\" odinger equation, which is a kind 
of the root of the Schr\" odinger equation, from the $d+1$ dimensional free 
Dirac equation. Starting from the Klein-Gordon equation and the Dirac equation 
in $d+1$ dimensions, the explicit transformation laws of the Schr\" odinger 
and the half Schr\" odinger fields under the Schr\" odinger symmetry transformation 
have been derived. 
We have also obtained the $3$ and $4$  dimensional super-Schr\" odinger algebra 
from the superconformal algebra in $4$ and $5$ dimensions. The algebra is realized 
by introducing two complex scalar and one (complex) spinor fields and the 
explicit transformation properties have been found. 
It could be interesting to find any system which has this kind of supersymmetry. 

\section*{Acknowledgments \label{Ack}}

This research has been supported by Global COE Program of Nagoya University (G07)
provided by the Ministry of Education, Culture, Sports, Science \&
Technology and by the JSPS Grant-in-Aid for Scientific Research (S)
\# 22224003 (SN).

\appendix

\section{Conformal transformations of scalar and spinor fields}

\subsection{Scalar}

Here we review on the conformal symmetry in $D$ dimensional space-time and 
show the invariance of the Klein-Gordon equation and the transformation law of the 
Klein-Gordon field under the conformal transformation. 
 
We consider the coordinate transformation:
\be
\label{I}
x^\mu = f^\mu\left( \tilde x^\nu \right)\, .
\ee
Under the coordinate transformation, the flat metric in $D$ dimensions
\be
\label{II}
ds^2 = \eta_{\mu\nu} dx^\mu dx^\nu\, ,
\ee
is changed as
\be
\label{III}
ds^2 = \eta_{\mu\nu} \frac{\partial f^\mu}{\partial \tilde x^\rho} 
\frac{\partial f^\nu}{\partial \tilde x^\sigma} 
d\tilde x^\rho d\tilde x^\sigma\, .
\ee
When the transformed metric is proportional to $\eta_{\mu\nu}$
\be
\label{IV}
\eta_{\rho\sigma} \frac{\partial f^\rho}{\partial \tilde x^\mu} 
\frac{\partial f^\sigma}{\partial \tilde x^\nu} 
= C\left( \tilde x^\tau \right) \eta_{\mu\nu}\, ,
\ee
the transformation (\ref{IV}) is called conformal transformation. 
Here $C\left( \tilde x^\tau \right)$ is a function of the coordinate and 
determined by multiplying Eq.~(\ref{IV}) 
with $\eta^{\mu\nu}$ as
\be
\label{V}
C\left( \tilde x^\tau \right) = \frac{1}{D} \eta^{\mu\nu} \eta_{\rho\sigma} 
\frac{\partial f^\rho}{\partial \tilde x^\mu} 
\frac{\partial f^\sigma}{\partial \tilde x^\nu} \, .
\ee
Then substituting (\ref{V}) into (\ref{IV}), we obtain
\be
\label{VI}
\eta_{\rho\sigma} \frac{\partial f^\rho}{\partial \tilde x^\mu} 
\frac{\partial f^\sigma}{\partial \tilde x^\nu} 
= \frac{1}{D} \eta_{\mu\nu} \eta_{\rho\sigma} \eta_{\alpha\beta} 
\frac{\partial f^\rho}{\partial \tilde x^\alpha} 
\frac{\partial f^\sigma}{\partial \tilde x^\beta}\, ,
\ee
We call $f^\mu (\tilde x^\nu)$ satisfying (\ref{VI}) as the conformal 
Killing vector. 
We now define a matrix $M^\mu_{\ \nu}$ by 
\be
\label{VII}
M^\mu_{\ \nu} \equiv \frac{\partial f^\mu \left( \tilde x^\rho \right)}
{\partial \tilde x^\nu}\, ,
\ee
Eq.~(\ref{VI}) can be rewritten as
\be
\label{VIII}
\eta_{\mu\nu} M^\mu_{\ \rho} M^\nu_{\ \sigma} 
= \frac{1}{D}\eta_{\rho\sigma} \eta^{\alpha\beta} \eta_{\mu\nu} 
M^\mu_{\ \alpha} M^\nu_{\ \beta}\, .
\ee
By regarding the quantity in (\ref{VIII}) as a matrix with indexes $\rho$ 
and $\sigma$ and considering the determinant, we obtain
\be
\label{IX}
\eta^{\alpha\beta} \eta_{\mu\nu} M^\mu_{\ \alpha} M^\nu_{\ \beta} 
= D M^{\frac{2}{D}}\, .
\ee
Here $M$ is the determinant of $M^\mu_{\ \nu}$: $M= \det M^\mu_{\ \nu}$. 

We now consider the coordinate transformation (\ref{I}) for the massless 
scalar field $\phi$: 
\be
\label{X}
S = \frac{1}{2} \int d^D x\, \eta^{\mu\nu} \partial_\mu \phi \partial_\nu \phi\, .
\ee
Then we obtain
\be
\label{X1}
S = \frac{1}{2} \int d^D \tilde x\, M \eta^{\mu\nu} \left( M^{-1} \right)_\mu^{\ \rho} 
\left( M^{-1} \right)_\nu^{\ \sigma}
\partial_\rho \left( M^{\frac{2-D}{2}} \tilde\phi \right) 
\partial_\sigma \left( M^{\frac{2-D}{2}} \tilde\phi \right) \, .
\ee
Here $\tilde \phi$ is defined by
\be
\label{X2}
\tilde \phi = M^{\frac{D-2}{2}} \phi\, .
\ee
By using (\ref{VIII}) and (\ref{IX}), we can rewrite the action (\ref{X1}) as
\be
\label{X3}
S = \frac{1}{2} \int d^D \tilde x \left\{ \eta^{\mu\nu} 
\partial_\mu \tilde\phi \partial_\nu \tilde\phi 
+ 2 \left(\frac{2-D}{2D} \right) \eta^{\mu\nu} \left( M^{-1} \partial_\mu M \right) 
\tilde \phi \partial_\nu \tilde \phi 
+ \left(\frac{2-D}{2D} \right)^2 \eta^{\mu\nu} \left( M^{-1} \partial_\mu M \right) 
\left( M^{-1} \partial_\nu M \right) \tilde \phi^2
\right\} \, .
\ee
By using the partial integration, we can further rewriting (\ref{X3}) as 
\be
\label{X4}
S = \frac{1}{2} \int d^D \tilde x \left\{ \eta^{\mu\nu} 
\partial_\mu \tilde\phi \partial_\nu \tilde\phi 
 - \frac{\left(D-2\right)}{4 \left(D-1\right) } \tilde \phi^2 
\left( - \left( D-1\right) \eta^{\mu\nu}\partial_\mu \partial_\nu \zeta 
 - \frac{\left(D-2\right) \left(D-1\right)}{4} \eta^{\mu\nu}
\partial_\mu \zeta \partial_\nu \zeta \right) 
\right\} \, .
\ee
Here $\zeta$ is defined by 
\be
\label{X5}
\zeta \equiv \frac{2}{D}\ln M \, .
\ee
Then Eqs.~(\ref{III}), (\ref{VI}), and (\ref{VII}) tell that 
the metric tensor $g_{\mu\nu}$ is given by
\be
\label{X6}
g_{\mu\nu} = \e^{\zeta} \eta_{\mu\nu}\, .
\ee
For the metric (\ref{X6}), the scalar curvature $R$ is given by
\be
\label{X7}
R = \left( - \left( D-1\right) \eta^{\mu\nu}\partial_\mu \partial_\nu \zeta 
 - \frac{\left( D-2\right) \left(D-1\right)}{4} \eta^{\mu\nu}
\partial_\mu \zeta \partial_\nu \zeta \right) \e^\zeta\, .
\ee
Since the metric (\ref{X6}) is obtained from the flat metric (\ref{II}), 
where the scalar curvature vanishes, the scalar curvature (\ref{X7}) 
also vanishes. 
Then the second term in (\ref{X4}) vanishes and we find
\be
\label{X8}
S = \frac{1}{2} \int d^D x\, \eta^{\mu\nu} 
\partial_\mu \phi \partial_\nu \phi 
= \frac{1}{2} \int d^D \tilde x\, \eta^{\mu\nu} 
\partial_\mu \tilde\phi \partial_\nu \tilde\phi \, ,
\ee
that is, the action (\ref{X}) of massless free scalar is invariant 
under the coordinate transformation (\ref{I}) and (\ref{X2}) where $f^\mu$ is 
the conformal Killing vector satisfying (\ref{VI}). 

\subsection{Spinor}

Here show the invariance of the Dirac equation and the transformation law of the 
Dirac field, that is, spinor field,  under the conformal transformation.

In order to investigate the conformal transformation in the spinor field, 
we consider the infinitesimal transformation
\be
\label{X9}
x^\mu = {\tilde x}^\mu + \delta x^\mu\, .
\ee
Then by defining,
\be
\label{X10}
m^\mu_{\ \nu} \equiv \frac{\partial \delta x^\mu}{\partial {\tilde x}^\nu}\, ,
\ee
we can rewrite the condition (\ref{VIII}) for the conformal transformation as 
\be
\label{X11}
\frac{1}{2}\left(m_{\mu\nu} + m_{\nu\mu} \right) = \frac{1}{D}\eta_{\mu\nu} m\, ,\quad 
m \equiv m^\mu_{\ \mu}\, .
\ee
The action of the Dirac fermion is given by
\be
\label{X12}
S = i \int d^D x \bar \psi \gamma^\mu \partial_\mu \psi\, .
\ee
We now consider the following infinitesimal transformation:
\be
\label{X16}
\psi = \left( 1 - \frac{D-1}{2D} + \frac{1}{2} m^{\mu\nu} \Sigma_{\mu\nu} \right) \tilde \psi\, ,
\ee
which gives 
\be
\label{X17}
\bar \psi = \left( 1 - \frac{D-1}{2D} - \frac{1}{2} m^{\mu\nu} \Sigma_{\mu\nu} \right) \tilde{\bar\psi}\, ,
\ee
Then we find 
\be
\label{X18}
S = i \int d^D x \left(\tilde{\bar\psi} \gamma^\mu \partial_\mu \tilde\psi 
 - m^\nu_{\ \mu}\tilde{\bar \psi} \gamma^\mu \partial_\nu \tilde\psi 
 - m \tilde{\bar\psi} \gamma^\mu \partial_\mu \tilde\psi 
 - \frac{D-1}{2D} \tilde{\bar \psi} \gamma^\mu \tilde\psi \partial_\mu m 
+ \frac{1}{2}  \tilde{\bar \psi} \gamma^\rho \Sigma_{\mu\nu} \tilde \psi \partial_\rho m^{\mu\nu} \right)\, . 
\ee
Since
\be
\label{X19}
d^D x \left( 1 - m \right) = d^D \tilde x \, ,\quad 
{\tilde \partial}_\mu  \equiv 
\frac{\partial}{\partial {\tilde x}_\mu} = \frac{\partial}{\partial x_\mu} 
 - m^\nu_{\ \mu} \frac{\partial}{\partial x_\nu}\, ,
\ee
we find
\be
\label{X20}
S = i \int d^D \tilde x \left(\tilde{\bar\psi} \gamma^\mu {\tilde \partial}_\mu \tilde\psi 
 - \frac{D-1}{2D} \tilde{\bar \psi} \gamma^\mu \tilde\psi {\tilde\partial}_\mu m 
+ \frac{1}{2}  \tilde{\bar \psi} \gamma^\rho \Sigma_{\mu\nu} \tilde\psi 
{\tilde \partial}_\rho m^{\mu\nu} \right)\, . 
\ee
In case of rotation (including the Lorentz transformation), translation, and dilatation, $m^{\mu\nu}$ is a 
constant or vanishes and therefore $\partial_\mu m = \partial_\rho m^{\mu\nu}$, which tells that the action 
(\ref{X12}) is invariant under the rotation (including the Lorentz transformation), translation, 
and dilatation. 
In case of the special conformal transformation with a infinitesimal parameter $c^\mu$, 
\be
\label{X21}
\delta x = c^\mu \left({\tilde x}\right)^2 - 2 {\tilde x}^\mu {\tilde x}^\rho c_\rho\, ,
\ee
we find 
\be
\label{X22}
\partial_\mu m = - 2 D c_\mu\, ,\quad 
\gamma^\rho \Sigma_{\mu\nu} \partial_\rho m^{\mu\nu} = - 2\left( D-1\right) c_\mu \gamma^\mu\, ,
\ee
and therefore the second and the third te7rms in (\ref{X18}) cancel with each other and the action 
(\ref{X12}) is invariant under the special conformal transformation (\ref{X19}).

\end{document}